\documentclass[12pt,a4paper]{article}

\usepackage{pifont}
\usepackage{listings}
\usepackage{xcolor}
\lstdefinelanguage{JavaScript}{
	morekeywords=[1]{break, continue, delete, else, for, function, if, in,
		new, return, this, typeof, var, void, while, with},
	morekeywords=[2]{false, null, true, boolean, number, undefined,
		Array, Boolean, Date, Math, Number, String, Object},
	morekeywords=[3]{eval, parseInt, parseFloat, escape, unescape},
	sensitive,
	morecomment=[s]{/*}{*/},
	morecomment=[l]//,
	morecomment=[s]{/**}{*/}, 
	morestring=[b]',
	morestring=[b]"
}[keywords, comments, strings]
\usepackage{subcaption}
\usepackage[margin=3pt,skip=6pt,belowskip=0pt,font={small,stretch=0.9},labelfont=bf]{caption}
\usepackage{amssymb}
\usepackage[abbreviations]{foreign}
\usepackage{hyperref}
\usepackage{ifthen}
\usepackage{changes}
\usepackage[export]{adjustbox}
\usepackage{eso-pic}
\usepackage[most]{tcolorbox}

\newcommand{\sys}{\textsc{CryptoAnalytics}\xspace}

\begin{document}
\def\confname{SoftwareX, Volume 26, May 2024}
\def\confyear{2024}
\def\confdoi{XXX}

\definecolor{yellowPaper}{HTML}{fff8ae}
\AddToShipoutPictureFG*{%
	\AtTextUpperLeft{%
		\adjustbox{raise=30pt}{
			\begin{tcolorbox}[width=1\textwidth,colback=yellowPaper,enhanced,frame hidden,sharp corners]  
				\centering\scriptsize
				\copyright~\confyear\ 	
				by Elsevier B.V. This is the author's version of the work.
				The final authenticated version is available online at \href{https://doi.org/10.1016/j.softx.2024.101663}{https://doi.org/10.1016/j.softx.2024.101663} 
				and has been published in  
				\confname.
		\end{tcolorbox}} 
	}%
}%

\renewcommand{\labelenumii}{\arabic{enumi}.\arabic{enumii}}

\title{\sys: Cryptocoins Price Forecasting with Machine Learning Techniques}

\author{Pasquale De Rosa \and Pascal Felber \and Valerio Schiavoni}
\date{University of Neuchâtel, Switzerland, first.last@unine.ch
}

\maketitle

\begin{abstract}
This paper introduces \sys, a software toolkit for cryptocoins price forecasting with machine learning (ML) techniques. 
Cryptocoins are tradable digital assets exchanged for specific trading prices. 
While history has shown the extreme volatility of such trading prices, the ability to efficiently model and forecast the time series resulting from the exchange price volatility remains an open research challenge.
Good results can been achieved with state-of-the-art ML techniques, including Gradient-Boosting Machines (GBMs) and Recurrent Neural Networks (RNNs). 
\sys is a software toolkit to easily train these models and make inference on up-to-date cryptocoin trading price data, with facilities to fetch datasets from one of the main leading aggregator websites, \ie CoinMarketCap, train models and infer the future trends. 
This software is implemented in Python.
It relies on PyTorch for the implementation of RNNs (LSTM and GRU), while for GBMs, it leverages on XgBoost, LightGBM and CatBoost. We follow an open science approach and release our code in the project \href{https://github.com/quapsale/cryptoanalytics-software}{repository}.
\end{abstract}

\section{Motivation and Significance}

\subsection{Introduction to \sys}

Cryptocoins are digitally-encrypted assets, used mostly in peer-to-peer networks. Depending on the underlying blockchain, cryptocoins are rewarded to nodes in the network.
History has shown the extreme volatility of cryptocoins trading prices. In the first place, one could consider these price trends \textit{unpredictable} and the resulting time series as a random walk. 
However, recent studies~\cite{1katsiampa, 2aslanidis} 
revealed the presence of co-movement among different coins and cross-correlation phenomena in cryptocoins market prices trends.

The main purpose of \sys is to provide third-party clients with a fast and reliable tool to leverage these co-movement patterns in order to forecast cryptocoins prices. 
\sys implements a wide range of state-of-the-art ML techniques for time series forecasting (LSTM \cite{3lstm}, GRU \cite{4gru}, XgBoost \cite{5xgboost}, LightGBM \cite{6lightgbm} or CatBoost \cite{7catboost}) in order to predict the market price of the desired cryptocurrency basing on other closely-related coins.

\subsection{Objective of \sys}
The goal of \sys is to provide a wide range of potential clients (like investors, institutions and/or goverments) with a reliable and easy-to-implement service to forecast cryptocoins prices. The crypto market is characterized by an extreme volatility, with sudden and continous changes in trading prices, as we will further discuss in \autoref{ssec:challenges}. With \sys it is possible to handle these complexities using an efficient and scalable tool. Moreover, we extend our analysis to the deployment of \sys into a production environment (\autoref{ssec:deployment}), making it possible to design a cryptocurrency prediction service designed to be adopted by either technical and non-technical users, with a very limited (or no) knowledge of ML algorithms.

\subsection{Scientific Contribution}

\sys has been used in the context of two scientific papers, presented respectively at the IFIP/DAIS 2022~\cite{8derosa} and at the ACM/DEBS 2023~\cite{9derosa} conferences. 
In \cite{8derosa}, \sys was leveraged to  investigate daily, weekly and monthly correlation patterns exhibited by the two main cryptocoins, Bitcoin and Ethereum, against a remaining set of 66 alt-coins.
Moreover, in \cite{9derosa}, \sys was used to study the trend correlations between and across a large set of 62 cryptocoins and subsequently to forecast Ethereum and Bitcoin price series basing on the trend of strongly-correlated crypto-assets. The results showed that \sys was able to provide reliable price forecasts with all the proposed state-of-the-art ML models.

\subsection{Theoretical Foundations}
To implement \sys, we considered two families of ML models: \emph{Gradient-Boosting Machines} (GBMs) and \emph{Recurrent Neural Networks} (RNNs), both adapted to cryptocoin price series forecasting. 
In \cite{9derosa} it was demonstrated that both GBMs and RNNs achieve reliable estimations of cryptocoin price series. More specifically, gradient-boosting machines were able to predict with high accuracy either \emph{stable} trends, with no trace of short-term peaks/falls, and unstable ones. 
On the other hand, recurrent neural networks were less accurate in modeling stable price trends. 
In the following paragraphs we describe these models in detail.

\subsubsection{Gradient-Boosting Machines}
\emph{GBMs} are "ensembles" of classification and regression trees \cite{9bisislr}. The main idea behind is to improve a single weak model by combining it with other weak models in order to generate a collective strong model. In GBMs, the iterative generation of weak models is determined minimizing the gradient over the chosen loss function. XGBoost, LightGBM and CatBoost are three notable state-of-the-art GBMs.

\emph{XGBoost} is an open-source, scalable and distributed GBM that builds trees in parallel rather than sequentially.

Microsoft's \emph{LightGBM} on the other hand is characterized by fast training speed and efficiency, fairly low memory usage and scalability.

\emph{CatBoost} introduces ordered boosting to avoid the prediction shift of the learned model, a common problem for traditional GBMs training.

\subsubsection{Recurrent Neural Networks}
\emph{RNNs} are a family of Neural Networks where the behavior of hidden neurons is not only determined by the activations in previous hidden layers, but also by the earlier stages \cite{9bisnn}. The training process of RNNs is usually complex, due to the unstable gradient problem: as a result of this phenomenon, vanilla RNNs are unable to model long term dependencies, lacking predictive ability when dealing with long sequences of data.
\emph{Gated} RNNs (LSTMs and GRUs) circumvent this problem in practical applications.

\emph{LSTM} embed \emph{cells} with internal recurrence (a self-loop), in addition to the outer recurrence of the RNN.

\emph{GRU} uses a single gating unit that simultaneously controls the forgetting factor and the decision to update the state unit.

\subsection{Cryptocurrency Market Challenges}
\label{ssec:challenges}
Cryptocurrencies surely represent a cutting-edge innovation in the field of financial technology. 
In \cite{10crypto}, cryptocoins are compared with two traditional and massively adopted financial assets: foreign exchange and stock. 
Basing on a four-year analysis of the daily close price trends, the authors conducted a comparatory study on five properties: volatility, centrality, clustering structure, robustness and risk. 
The cryptocurrency market proved to be more similar to that of the stock one, but characterized by an higher rate of fragility and risk.
Evidence of this fragility can be found in the extreme volatility of cryptocoin trading prices, and possible reasons for this behavior include lack of adequate regulation, the inherent speculative nature of these assets, the lack of an institutional guarantor and \emph{pump-and-dump} actions enacted by large stakes (\ie, \emph{whales} owning large percentages of the issued coins).
The complex nature of this market, characterized by sudden and several price fluctuations, determines the need for a reliable guidance for all the potential investors. \sys can represent an ideal solution to address these open challenges, by providing a fast, easy-to-implement and reliable solution to forecast future cryptocoin prices basing on past trend correlations between these crypto-assets.

\subsection{Software Requirements}

There are no minimum hardware requirements for \sys. 
Since the training process for RNNs can be computationally intensive, \sys automatically detects any CUDA-capable \cite{10cuda} hardware accelerator (\ie GPUs) present on the machine, and uses it if available. 
Otherwise, the training process is performed on the CPU. 
Python v3.9.10 is required to run our tool.

\section{Software Description}
This section provides an overview of the \sys software architecture with a short description of its functionalities.

\begin{figure}[t]
	\centering
	\includegraphics[scale=0.7]{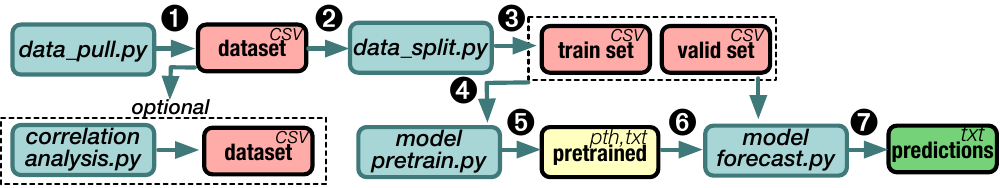}
	\caption{Workflow and functionalities of \sys.}
	\label{fig:workflow}
\end{figure}

\subsection{Software Architecture}

\sys implements a cryptocoin forecasting pipeline, driven by easy-to-use processes, for which we implement a command line interface over five major functionalities: \emph{(1)} data pull, \emph{(2)} data split, \emph{(3)} model pretrain, \emph{(4)} model forecast, and optionally \emph{(5)} correlation analysis.
The execution workflow is depicted in Figure~\ref{fig:workflow} and the software functionalities are described in detail next.

\subsection{Software Functionalities}

The \textit{data pull} functionality provides methods to generate a new dataset of Open-High-Low-Close (OHLC) cryptocoin prices from the leading aggregator CoinMarketCap \cite{11coinmarketcap}.
OHLC charts are commonly used to illustrate the price movements of financial instruments (in this case cryptocurrencies) over time. 
Upon execution (Fig.~\ref{fig:workflow}-\ding{202}), \textit{data pull} produces a new dataset ready to be used in the further analyses.
User-specified arguments are: the destination directory, the file name, the list of coins to include in the dataset (in \texttt{.json} format), start and end date of the pull (in \%d-\%m-\%Y format).
This dataset is sent as input (Fig.~\ref{fig:workflow}-\ding{203}) to the following step.

The \textit{data split} functionality provides methods to generate train and validation sets from the original data.
Upon execution, \textit{data split} produces two dataset splits in .csv format: train and validation (Fig.~\ref{fig:workflow}-\ding{204}).
Train and validation sets are used in the model training process (Fig.~\ref{fig:workflow}-\ding{205}). Moreover, the validation set is adopted in the forecasting phase as well to predict the feature coins price series for the forecasting horizon. 
User-specified arguments are: the destination directory, the file names, the path to the dataset gathered before, the price variable to consider (either the average OHLC price or just the Close price) and the ratios for train/validation split (as floats).

The \textit{model pretrain} functionality provides methods to efficiently pretrain ML models (RNNs and GBMs) for cryptocoins price forecast. The execution of this command (Fig.~\ref{fig:workflow}-\ding{206}) produces a pretrained model that is used further (Fig.~\ref{fig:workflow}-\ding{207}) in the final forecasting phase. More specifically, pretrained RNNs are stored as .pth, extensions designed to store serialized PyTorch state dictionaries, while pretrained GBMs are stored as .txt files. 
For all the considered ML models, the price forecast for a given user-specified cryptocurrency is given by observing the fluctuation of other highly-correlated coin series (namely, these will be the "feature variables"). In order to pre-select these feature coins, the user might conduct a preliminary correlation analysis (detailed further).
For the experimental setting, a list of user-specified configurations is needed. These configurations vary with respect to the ML model adopted. For RNNs, it has to be declared the size of the network (number of hidden layers and neurons), the number of training epochs and the batch size (namely, the number of samples processed before a model update). For GBMs, the user has to specify the number of tree splits generated by the model. Common configuration to all considered models are: the random seeds, the learning rate and the patience (\ie number of "tolerated" consecutive epochs/tree splits without a model improvement before early-stopping the training).
User-specified arguments are: the destination directory, the file name, the path to the train and validation sets, the ML model to use (either LSTM, GRU, XgBoost, LightGBM or CatBoost), the target coin to predict, the list of coins to use as feature/predicting variables (in .json format) and the list of configurations to use for the experimental setting (in .json format).

The \textit{model forecast} functionality provides methods to predict future cryptocoins prices using the pretrained. The validation set is used to fit a Holt-Winters Exponential Smoothing model \cite{12holtwinters} to forecast the feature coin price series for the user-specified predicting horizon. The predicted feature coins will be then fed to the pretrained ML model to generate price predictions for the selected target cryptocurrency. Moreover,   
the resulting predictions are stored in .txt format (Fig.~\ref{fig:workflow}-\ding{208}). 
User-specified arguments are: the destination directory, the file name, the forecasting horizon (number of future daily prices to predict), the path to the validation set, the path to the pretrained model, the ML model to use, the target coin to predict and the list of coins to use as feature/predicting variables (in .json format). These arguments must be the same used for the pretraining process.

Finally, the \textit{correlation analysis} functionality provides methods to analyze correlations among cryptocoin prices. 
It is not required for the main price prediction flow, but can be useful to pre-select highly correlated cryptocoins to use as feature variables.
User-specified arguments are: the destination directory, the file names, the path to the original dataset, the price variable to consider (either the average OHLC price or the Close price), the time window to use for computations  (either daily, weekly or monthly) and the correlation method to use (either Pearson, Kendall or Spearman \cite{13correlation}).

\section{Illustrative example}

This section provides a comprehensive illustrative example of the full price prediction flow with the \sys toolkit. File names and directories are listed as they appear in the project GitHub \href{https://github.com/quapsale/cryptoanalytics-software}{repository}. 

\subsection{Data Pull}

We start by pulling the dataset of cryptocoin OHLC market prices from CoinMarketCap. 
We decide to pull data for a list of 8 cryptocoins (BTC, ETH, USDT, USDC, XRP, BUSD, ADA, DOGE), defined in the "coins.json" file inside the /examples directory. 
Moreover, we adopt a time frame of three months (15-08-2023 to 15-11-2023). 
The command used is listed below:

\begin{lstlisting}[language=Bash,
	xleftmargin=65pt,
	xrightmargin=65pt,
	label={lst:bash-data-pull},
	caption=Data pull command.]
	python data_pull.py \
	--filename "dataset" \
	--coins "examples/coins.json" \
	--start "15-08-2023" \
	--end "15-11-2023"	
\end{lstlisting}

This command generates a "dataset.csv" file in the current working directory. This dataset is made up by 6 columns (Date, Open, High, Low, Close and Coin).

\subsection{Data Split}

We then proceed by splitting the pulled data into train and validation sets. In the example we use a conventional split ratio of train = 80\% and valid = 20\%. The price variable considered for predictions is the average OHLC price.
The command used is listed below:

\begin{lstlisting}[language=Bash,
	xleftmargin=65pt,
	xrightmargin=65pt,
	label={lst:bash-data-split},
	caption=Data split command.]
	python data_split.py \
	--filenames "train" "valid" \
	--data "dataset.csv" \
	--variable "avg_ohlc" \
	--train (*@\textcolor{red}{0.8}@*) \
	--valid (*@\textcolor{red}{0.2}@*)
\end{lstlisting}

This command generates two files "train.csv" and "valid.csv" in the current working directory. The train dataset contains observation ranging from 15-08-2023 to 28-10-2023. The validation dataset contains observation ranging from 29-10-2023 to 15-11-2023.

\subsection{Correlation Analysis (optional)}

The correlation analysis is optional, but can be useful to pre-select cryptocoins to use as feature variables for the model pretrain and forecast. 
In our example, we decide to use Bitcoin (BTC) as the target coin to predict and further identify a set of 5 highly correlated ($\textrm{Pearson} > 0.5$) coins to BTC. We use the Pearson coefficient to compute these correlations over a daily sliding window.
The command used is listed below:

\begin{lstlisting}[language=Bash,
	xleftmargin=65pt,
	xrightmargin=65pt,
	label={lst:bash-corr-analysis},
	caption=Correlation analysis command.]
	python correlation_analysis.py \
	--filename "correlations" \
	--data "dataset.csv" \
	--window "daily" \
	--method "pearson"
\end{lstlisting}

This command generates a "correlations.csv" file in the current working directory, containing a cross-correlation table with all the cryptocoins in the dataset. We exclude the \textit{stablecoins} and pre-select ETH, XRP, DOGE and ADA as feature variables. 

\subsection{Model Pretrain}

We can then pretrain our ML model to forecast the average OHLC price of Bitcoin.
In this demonstration, we choose to train a LSTM (Long-Short Term Memory) neural network with a set of predefined configurations, specified in the "config\_nn.json" file inside the /examples directory.
In the same path we defined the set of pre-selected feature coins inside the "features.json" file. 
The command used is listed below:

\begin{lstlisting}[language=Bash,
	xleftmargin=65pt,
	xrightmargin=65pt,
	label={lst:bash-mod-pretrain},
	caption=Model pretrain command.]
	python model_pretrain.py \
	--filename "lstm" \
	--train "train.csv" \
	--valid "valid.csv" \
	--target "btc" \
	--features "examples/features.json" \
	--model "lstm" \
	--config "examples/config_nn.json"
\end{lstlisting}

This command generates a "lstm.pth" file in the current working directory, containing the pretrained LSTM model.

\subsection{Model Forecast}

Finally, we can use the pretrained model to make inference on the unseen data.
Our aim is to predict the price series of Bitcoin for the following week (namely, the forecasting horizon will be of 7 days).
The configurations to use in this phase must be the same adopted for pretraining.
The command used is listed below:

\begin{lstlisting}[language=Bash,
	xleftmargin=65pt,
	xrightmargin=65pt,
	label={lst:bash-mod-forecast},
	caption=Model forecast command.]
	python model_forecast.py \
	--filename "predictions" \
	(*@\textcolor{black}{--valid}@*) "valid.csv" \
	--horizon (*@\textcolor{red}{7}@*) \
	--pretrained "lstm.pth" \
	--target "btc" \
	--features "examples/features.json" \
	--model "lstm"
\end{lstlisting}

This command generates the final output of the price prediction flow, that is a "predictions.txt" file containing the forecasts of Bitcoin average OHLC prices for the weekly time horizon (16-11-2023 to 22-11-2023).

The resulting predictions showed a fairly good accuracy, with a Mean Absolute Percentage Error  (MAPE) $\approx 6.57\%$\footnote{According to \cite{19mape}, a MAPE $<10\%$ corresponds to an highly accurate forecasting model.} and a Root Mean Squared Error (RMSE) $\approx 2438.37$ USD respect to the observed average OHLC prices from CoinMarketCap.

\begin{lstlisting}[language=Bash,
	xleftmargin=65pt,
	xrightmargin=65pt,
	label={lst:json-forecast},
	caption=Bitcoin weekly price predictions vs real (in USD) as of 15-11-2023.]
	Predicted: 34368.3, Real: 36878.7
	Predicted: 34363.8, Real: 36341.8
	Predicted: 34366.0, Real: 36570.9 
	Predicted: 34368.5, Real: 36974.1
	Predicted: 34365.5, Real: 37372.6
	Predicted: 34368.4, Real: 36682.0
	Predicted: 34364.3, Real: 36679.2
\end{lstlisting}

\section{Software Deployment}
\label{ssec:deployment}
In this section, we discuss how to deploy \sys to build fast and reliable cryptocurrency prediction services with ML algorithms. 
To conduct our analysis, we analyzed the performance of \sys using three benchmark frameworks: TorchServe \cite{14ts}, BentoML \cite{15bentoml} and MLFlow \cite{16mlflow}. Moreover, we considered two MLFlow scenarios: a base one, with local Flask server, and a second one with MLServer \cite{17mlserver}.

We performed our benchmark study on Ubuntu 22.04.2 LTS, Linux kernel 5.15.0-88-generic, 64-core CPU Intel\texttrademark\ Xeon\texttrademark\ E5-2683 v4 clocked at 2.10GHz with 128 GB RAM.

\begin{figure}[t]
	\centering
	\includegraphics[scale=0.7]{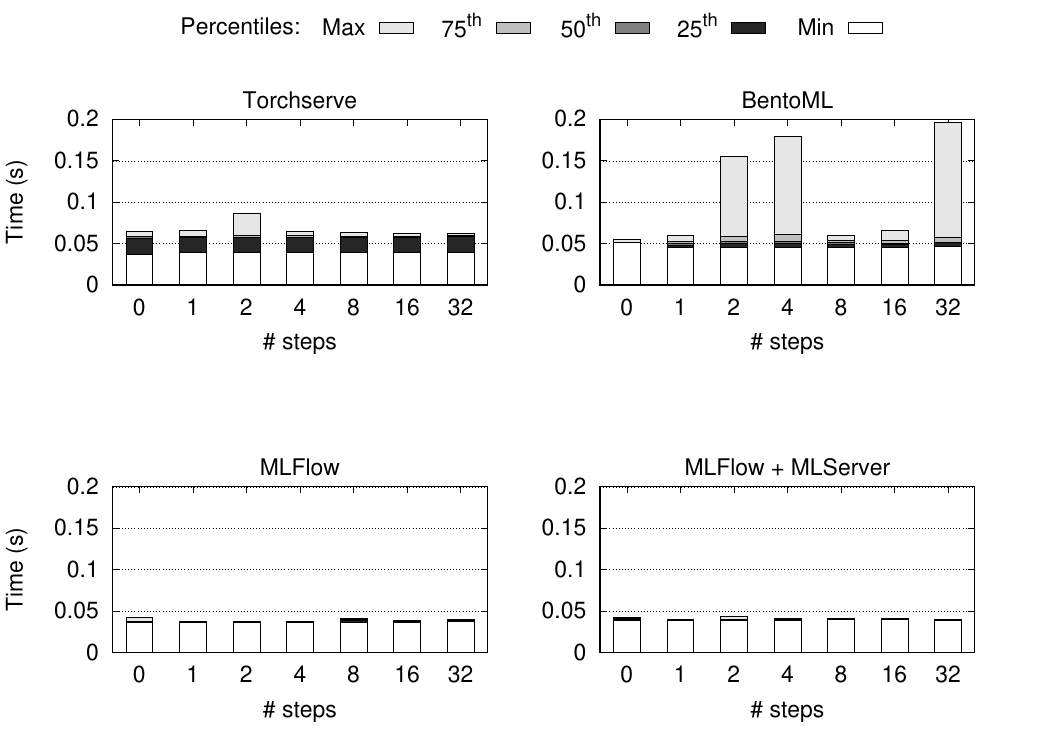}
	\caption{Benchmark of deployment frameworks for \sys.}
	\label{fig:cryptoserve}
\end{figure}

For the evaluation, we used oha \cite{18oha} to submit multiple prediction requests to the \sys server for a total time of 2 minutes. 
We analyzed the response time in relation to the number of steps (future daily prices to predict), ranging from 0 (the baseline) to 32. 
The results are shown in \autoref{fig:cryptoserve}. 

Overall, there is no significant impact of the number of steps on the response time. 
MLFlow shows the best performance, with almost all response delays distributing around the minimum value of $\approx 0.04$ seconds. These results are consistent in both MLFlow scenarios.

Examples of ready-to-use \sys deployments for each framework are available in the project  \href{https://github.com/quapsale/cryptoanalytics-software/tree/main/deployment}{repository}. 

\section{Impact}
The analysis and forecasting of cryptocoins price trends still offers challenges in both academic and industrial contexts, due to the abrupt nature of their fluctuations over time. 
The rapid development of ML techniques allowed new approaches to time series modeling, as a valid alternative to traditional econometrics ones (\eg, ARMA and VAR processes~\cite{20hamilton}). 

However, these algorithms rely on complex statistical and mathematical concepts, making it difficult for end users lacking prior technical knowledge to train and validate these models. 
Moreover, also skilled users could benefit of a time-saving solution that does not require any experimental setup, but just a simple command-line interface that embeds the entire ML workflow. 

\sys aims to be a solution for either types of users, providing them with a simple, clear and effective interface that allows to obtain cryptocoins prices predictions with just few lines of bash code. 

Moreover, we showed that \sys can be deployed efficiently using state-of-the art frameworks to build fast and reliable cryptocoins prediction services for a wide range of users.  

The nature of this software makes it a good fit also for business-oriented applications.
Indeed, fluctuations in the prices of digital currencies naturally lead investors to have concerns, making the behaviour of crypto markets very difficult to predict. For this reason, a fast, easy-to-implement and reliable solution like \sys can allow to accurately forecast cryptocoin prices, in order to not only help investors in making decisions, but also governments in designing regulatory policies \cite{21crypto}.
As we previously mentioned, the use of \sys is attested in \cite{8derosa} and \cite{9derosa}, where it has been applied to the analysis of cross-correlations patterns over a large set of cryptocoins in order to forecast their future price series.

\section{Related Work}

\subsection{Cryptocoin Prediction Tools}
Since cryptocurrencies became significant investment drivers in the recent years, the ability to predict their prices is crucial for the investors to make informed decisions. Here we will describe briefly some of the most well-known tools on the online market that represent possible alternatives to \sys.
\begin{enumerate}
	\item WalletInvestor \cite{22walletinvestor} is an online prediction site that makes use of ML algorithms to produce price forecasts. WalletInvestor's cryptocurrency predictions are based on multiple economic factors like: changes in the exchange rates, trade volumes, volatilities of the past period. Long term (3 months, 1 year and 5 year) forecasts for more than 800 coins are available for free.
	\item CryptoPredictions \cite{23cryptopredictions} is an online tool that makes use of historical exchange rates and market data to predict the future price trend of a given coin. The forecasting algorithm uses a combination of linear and polynomial regressions. Daily, monthly and yearly predictions for over 8000 cryptocurrencies are available for free.
	\item DigitalCoinPrice \cite{24digitalcoinprice} is a price-tracking website for cryptocurrencies. It provides as well a price prediction service for the listed cryptocoins based on historical data. Monthly and yearly predictions for over 8000 cryptocurrencies are available for free.
	\item CryptoRating \cite{25cryptorating} is an online service for cryptocoin forecast and analysis. It makes use of a unique and elaborate ML algorithm that takes into account multiple factors, such as the Crypto Volatility Index. A paid subscription is required to get the full access to daily, monthly and yearly predictions for 100 cryptocurrencies.

\end{enumerate}

\subsection{Innovation and Advancements of \sys}
Compared to other solutions already present on the market, \sys has surely the advantage of being an open-source, free and easy to access software. This aspect makes \sys a useful tool either for industrial and research applications. Moreover, the publicity of the code makes possible for other researchers and/or interested users to contribute and expand its scope. This can involve either the data sources (at the current state, just CoinMarketCap) and the ML algorithms to pretrain for practical forecasting. Also, the pool of models managed by \sys goes beyond traditional ML and time series modeling algorithms, involving Deep Learning ones (like RNNs).

A noticeable difference with the current state-of-the-art is that \sys provides individual price forecasts for each coin by looking at the time series of other highly-correlated cryptocurrencies, instead of the historical (lagged) data of the target to predict.
This approach, as anticipated in the previous section, was successfully adopted in \cite{9derosa}. Moreover, the study of co-movement and cross-correlation events in cryptocurrency market trends has been widely explored in the recent literature.
In \cite{1katsiampa}, the author found evidence of interdependencies between the Bitcoin and Ether, with price responsiveness to major news in the market.
In \cite{2aslanidis}, authors first showed that cryptocurrencies exhibit similar mean correlation among them, and then detected an independent behavior respect to other financial markets.

\section{Conclusions}
Cryptocoins show very volatile trends.
Despite this behavior, the presence of co-movement and cross-correlation patterns among cryptocoins suggests that it might be possible to forecast a coin price evolution by observing fluctuations in other coins' trends. 
Machine Learning (ML) techniques like GBMs and RNNs constitute nowadays the state-of-the-art in modeling and predicting complex, time-varying and large-scale price series.

We presented \sys, a toolkit based on Python, designed to easily train these models and make inference on up-to-date cryptocoin price data.
Moreover, we discussed how to deploy \sys to build fast and reliable cryptocurrency prediction services using state-of-the-art frameworks (TorchServe, BentoML and MLFlow).
   
\sys can represent a useful tool, either in business or academic applications, to gather information from cryptocoins and leverage their co-movement behaviors in order to model and forecast the trends of the asset prices.

\end{document}